\documentclass{aa}  
\usepackage{graphicx}
 \usepackage[latin1]{inputenc}
 \usepackage{natbib}
\bibpunct{(}{)}{;}{a}{}{,} 
\def\kms{~km~s$^{-1}$}
\def\cmmt{~cm$^{-3}$}
\def\cmmd{~cm$^{-2}$}

\def\le{$\leq$}

\def\deg{^\circ}

\begin{document}
   \title{Coupling the dynamics and the molecular chemistry in the Galactic center}


   \author{N. J.  Rodriguez-Fernandez  \inst{1,2}   
          \and F. Combes \inst{3}
           \and J. Martin-Pintado\inst{4} 
            \and T. L. Wilson\inst{5}
            \and A. Apponi\inst{6}
     }

   \offprints{N. J. Rodriguez-Fernandez}

   \institute{Observatoire de Bordeaux, L3AB (UMR 5804)/OASU, CNRS/Universit\'e Bordeaux 1, 
    BP 89, 2 rue de l'Observatoire, 33270 Floirac, France \\
    \email{nemesio.rodriguez@obs.u-bordeaux1.fr}
    \and Universit\'e Denis Diderot (Paris VII) \& Observatoire de Paris,  61 Av de  l'Observatoire, 75014 Paris, France
    \and LERMA,  Observatoire de Paris,  61 Av de  l'Observatoire, 75014 Paris, France
            \and DAMIR, IEM, CSIC, Serrano 121, Madrid, Spain     
            \and ESO,  Karl-Schwarzschild-Str. 2, D-85748 Garching bei M\"unchen, Germany
            \and Steward Observatory, University of Arizona, Tucson, AZ 85721, USA    }

   \date{Received September 15, 1996; accepted March 16, 1997}

 
\authorrunning{Rodriguez-Fernandez et al.}
\titlerunning{Dynamics and molecular chemistry in the Galactic center}

  \abstract
   {Most of the  Galactic center (GC) gas moves in nearly circular orbits in a nuclear ring (hereafter Galactic center ring, GCR).  This  is the case of cloud complexes such as Sgr A or Sgr B, where the gas is dense, warm and exhibits a rich molecular chemistry. The origin of these properties is thought to be some kind of shock, in particular due to the large scale dynamics of the Galaxy.  In addition, there are gas  clouds moving in highly non-circular orbits known by observations of low density tracers as CO(1-0). The physical conditions of the clouds  moving with non-circular velocities are not well-known.}
   {We have studied the physical conditions of the gas in non-circular orbits with the aim of  better understanding the origin of the outstanding physical  conditions of the GC molecular gas and the possible effect of the large scale dynamics on these physical conditions.}
   {Using published CO(1-0) data, we have selected a set of clouds belonging to all the kinematical components seen in the longitude-velocity  diagram of the GC. We have done a survey of dense gas in all the components  using the $J=2-1$ lines of CS and SiO as tracers of high density gas and shock chemistry. }
   {We have detected CS and SiO emission in all the kinematical components. The gas density and the SiO abundance of the clouds in  non-circular orbits are similar those  in the GCR. 
Therefore, in all the kinematical components there are dense clouds that can withstand the tidal shear. However, there is no evidence of star formation  outside the GCR.  The high relative velocity and shear expected in the dust-lanes  along the bar major axis could inhibit the star formation process, as observed in other galaxies.  The high SiO abundances derived in the non-circular velocity clouds are likely  due to the large-scale shocks that created the dust lanes.}
   {}


\maketitle

%

\section{Introduction}
\begin{figure}[tbhp] %
\begin{center}
\includegraphics[width=8.5cm]{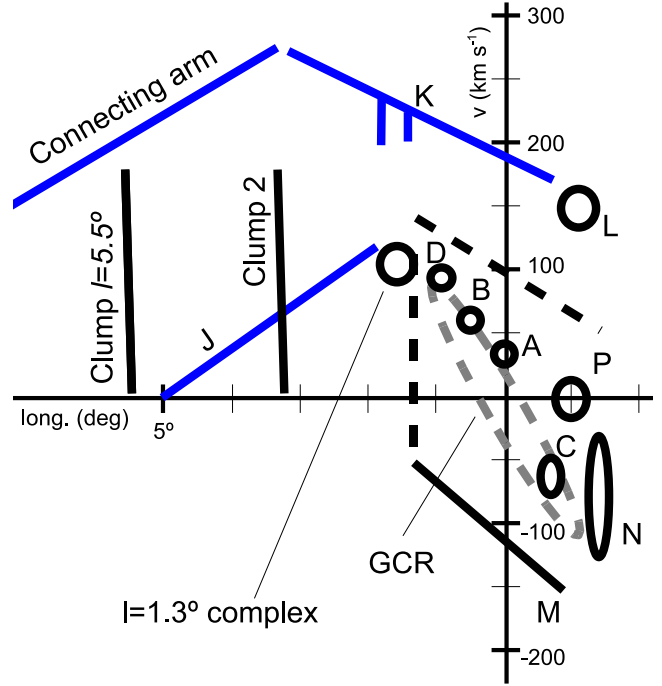}
\caption{Schema illustrating the main kinematical structures in the longitude-velocity diagram of the inner degrees of the Galaxy. The labels A to D stand for Sgr A to Sgr D complexes. See the text for an explanation of the other labels}
\label{fig_lv}
\end{center}
\end{figure}

\begin{figure}[tbhp] %
\begin{center}
\includegraphics[width=6cm]{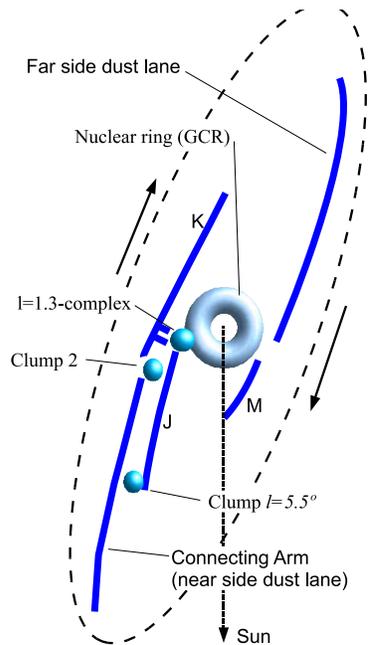}
\caption{Schema of a plausible face on view of the Galactic center}
\label{fig_face_on}
\end{center}
\end{figure}

The longitude-latitude maps of the Galactic center (GC) show several major cloud complexes: the $l=1.3\deg$-complex (a huge molecular complex  located at longitudes between $l=1.2^\circ$ and $l=1.6 \deg$), the Sgr D complex at $l=1\deg$, the Sgr B complex near $l=0.7\deg$, the Sgr A cloud near $l=0\deg$ and the  Sgr C cloud near $l=-0.5\deg$ \citep[see for instance,][]{Scoville72,Liszt78,Bania77,Bally87,  Dahmen97, Oka98a, Oka98b, Bitran97}. 
In the longitude-velocity diagram (hereafter \emph{lv}-diagram, see Fig. \ref{fig_lv}), the Sgr A ... D complexes  are approximately located along a line passing through the origin as expected for gas moving along an almost circular ring. Therefore, in a face-on view of the Galactic center these complexes would be located in a circular ring (Fig. \ref{fig_face_on}). \cite{Sofue95} proposed that this ring is composed of two  arms (Arm I and Arm II). Hereafter, we refer to this structure as the Galactic center ring (GCR). However, the GCR clouds and the $l=1.3\deg$-complex are not the only cloud components in the GC.  The $lv$-diagram also exhibits many structures corresponding to clouds  moving in non-circular trajectories like the Clump 2, the Connecting Arm, or the structures that we have labeled as J, K, L, M, N, O and P (Fig. \ref{fig_lv}). The easiest explanation for the non-circular trajectories is the gas response to a bar potential \citep{Binney91,  Fux99}.

In the inner region of a barred galaxy there are two main families of closed orbits in the reference frame of the bar \citep[][]{Contopoulos80}: the $x_1$ orbits, which are elongated along the bar major axis and the $x_2$ orbits, which are elongated along the minor axis of the bar, when there exists inner Linblad resonances (ILR). Numerical simulations of  the gas dynamics \citep{Athanassoula92,  Fux99} show that the gas in the center tends to follow nearly circular $x_2$ orbits creating a nuclear ring (the GCR). At larger radii, beyond the ILRs, the gas tends to follow $x_1$ orbits along the bar. In fact, the gas clouds are subject to collisions, and do not follow strictly these orbits, but are frequently kicked off into neighboring orbits, in general with lower energy. In particular, at the transition region between the $x_1$ and $x_2$ flows, collisions are favored, and there might be a scatter, or ``spray'' of gas clouds moving along the $x_1$ flow.
The sprayed gas will collide with the material on the opposite side of the bar major axis giving rise
to  shocks  in the molecular cloud component   and to the characteristic dust lanes that are observed near the leading edges of the bars in external galaxies.
Losing energy and angular momentum, the gas  of the dust lanes will progressively spiral down to the nuclear ring and it will fall in  after only a few  rotations   \citep[see Fig. 20 of][]{Fux99}.

In the GC, \cite{Fux99} has identified the Connecting Arm as the near-side dust lane.  The structure  K, which  seems to be linked to the Connecting Arm at $l \sim 3\deg$ (Fig.~\ref{fig_lv}),  represents gas clouds that move in the  decelerating section of elongated  orbits, i.e, gas that does not fall in the nuclear ring at the first passage around the $x_1-x_2$ interaction region (see Fig.~\ref{fig_face_on}).   The structure J  has nearly the same inclination as that of the Connecting Arm (Fig.~\ref{fig_lv}), and it could be a kind of  small dust lane, but in contrast with the Connecting Arm and the structure K, the gas in the structure J will easily interact with the material of the nuclear disk in the area of the $l=1.3\deg$-complex  (see Fig. \ref{fig_face_on}).  Thus, the $l=1.3\deg$-complex would be the contact point of the dust lane and the  GCR as proposed by \cite{Huttemeister98} and \cite{Fux99}.  The clump at $l=5\deg$ and the Clump 2 are gas clouds that are about to enter the dust lane shock. The $lv$-diagram at  $l<0$ is more difficult to interpret. One should note that the  $lv$-diagram is  not expected to be symmetric. Due to perspective effects  the far side dust lane is elongated along the line of sight and thus it is  a vertical structure in the $lv$-diagram close to the extremity of the nuclear ring \citep[see Fig. 23 of][]{Fux99}.  
The complexes L, N, and P could indeed belong to the far-side dust lane shock. 
The structure M  would be   part of the gas spray due to the interaction of the far-side dust lane and the GCR.

The physical conditions of the neutral gas in the GC  have been studied by  observing high density tracers like CS \citep{Bally88, Tsuboi99} or HCN \citep{Jackson96} and by multi-level studies of NH$_3$ \citep{Huttemeister98},  H$_2$ \citep{Rodriguez01, Rodriguez04} or CO   \citep{Sawada01, Martin04}. These works, among others, have shown that the clouds in the GCR and  the $l=1.3\deg$-complex are dense ($10^4$ \cmmt) and warm (up to $\sim 150$ K). 
In addition, the GCR clouds present widespread emission and high abundances of molecules like SiO, SO, CH$_3$OH, or C$_2$H$_5$OH  \citep{Martin-Pintado97, Martin-Pintado01, Lis01, Huttemeister98}, which are considered good shock tracers.
Indeed, shocks are commonly invoked to explain the molecular chemistry and the high temperatures of the neutral gas in the GCR, the $l=1.3\deg$-complex and the Clump 2  \citep{Martin-Pintado97,  Martin-Pintado01, Huttemeister93, Huttemeister98, Lis01, Rodriguez04}. The origin of the shocks could be related to Wolf-Rayet winds \citep{Martin-Pintado99} or to cloud-cloud collisions  induced by the large scale dynamics of the Galaxy \citep{Wilson82, Huttemeister98, Martin-Pintado01, Sato00, Rodriguez00, Rodriguez04}.

In contrast to the GCR and the $l=1.3\deg$-complex, little is known about the physical conditions of the GC  clouds with non-circular velocities. 
Up to now, the  only large scale maps that have been used to study the physical conditions of the ensemble of the GC molecular gas are those of  \cite{Oka98a} and \cite{Sawada01}. They have mapped the CO(2-1) and CO(1-0) lines and found a rather constant line ratio  ($\sim1$) in the  GC clouds (including Clump 2). 
A few clouds of the upper-right corner of the $lv$-diagram have been  observed in SiO and NH$_3$ by \cite{Huttemeister93,Huttemeister98} and in H$_2$ and fine structure lines by \cite{Rodriguez04}. Their densities, temperatures and excitation mechanisms seem to be similar to the GCR clouds. 

Thus, in contrast to the GCR cloud complexes, the physical conditions of the gas with non-circular velocities that constitute the characteristic shape of the  $lv$-diagram are not well-known. However, the study of the properties of all these kinematical components can give us important  hints on the origin of the outstanding physical conditions of the GC clouds and on the possible links between these  physical conditions  and  the global gas dynamics. 
This is the purpose of this paper.
We have selected a sample of clouds belonging to all the kinematic components seen in the GC $lv$-diagram and we have undertaken a survey of dense gas using CS and SiO as tracers. Due to their  high dipole moments, both CS and SiO are high density tracers ($\geq 10^4$ cm$^{-3}$). In addition, the SiO molecule traces dense gas affected by shocks \citep{Martin-Pintado92}. Since the CS and SiO molecules are quite similar and they have similar energy level distributions, the SiO/CS line ratio is expected to depend mainly on the relative abundance of the two molecules.

\section{Observations}

We have selected  61 positions  in the longitude range $3.5^\circ>l>-1.5^\circ$ using the \cite{Bally87}  CO(1-0) data cube. In fact, due to the crowded velocity fields in the GC, with these  61 pointings we have been able to study 161 clouds  (both GCR and non-GCR clouds). Table \ref{tab_results}  presents the Galactic coordinates and velocities of the different clouds.  For an easy presentation and discussion of the results  we have grouped the clouds that occupy a same region in the $(l, b, v)$ space or that  belong to one of the coherent kinematical structures discussed above.  

We have observed  43 of the selected positions in the J=2-1 line of CS and SiO using the University of Arizona 12 meter antenna at Kitt Peak in February 2003. The half power beam width of the telescope is 68'' and 77''  at the frequencies of the CS and SiO lines,  respectively.  
The temperatures, in the $T_R^*$ scale, have been converted to main-beam temperatures ($T_{mb}$) using  the expression $T_{mb} =  T_R^*  / \eta_m^* $  with  corrected main beam efficiencies $\eta_m^*$  of 0.95 and 0.94 for the SiO and CS spectra, respectively.  As backends, we used filter-banks with 1 MHz channels, which provide a velocity resolution of 3.06 and 3.45 \kms~ for the CS and SiO lines, respectively. The $rms$ noise in 1 MHz channels is in the range $0.04-0.11$ K for the CS spectra and in the range $28-51$ mK for the SiO spectra.

 The other 18 positions  were observed with the  IRAM 30 meter telescope in July 2003. The two 100 GHz receivers were used to simultaneously observe the two lines. The 30m telescope  beam size at the frequencies of the CS and SiO  lines is 25.5'' and 29'' respectively. The temperatures, in the  $T_A^*$ scale, have been  converted to main beam temperatures using the expression $T_{mb}=(F_{eff}/B_{eff}) \, T_A^*$ with a forward efficiency ($F_{eff}$) of 0.95 and beam efficiencies ($B_{eff}$) of 0.78 and 0.77 for the SiO and CS spectra, respectively.    As backends, we also used filter-banks with 1 MHz channels. 
The $rms$ noise in 1 MHz channels vary from source to source in the range $15-75$ mK for the CS spectra and  in the range $11-57$ mK for the SiO spectra. The lowest $rms$ are needed to obtain good detections of some  clouds with non-circular velocities as those in the structure K. 
For comparison,  the large scale CS(2-1) map of  \cite{Bally87}  had a  $rms$  noise of  0.15 K in 1MHz filters.

We have also observed the $^{13}$CO (2-1) and (1-0) lines towards 12 of the  positions observed with the 30m telescope (see Table \ref{tab_lvg}).
 The IRAM 30m beam size at the  frequency of these lines is $11^{''}$ and $22^{''}$, respectively. The temperatures have been converted to $T_{mb}$ using  forward  and beam efficiencies of 0.95 and 0.75 for the $^{13}$CO(1-0) line and 0.91 and 0.55 for the  $^{13}$CO(2-1) line.  All the observations were done in position switching mode. The emission free OFF positions have been carefully selected from the  \cite{Oka96} maps and our own spectra of different observational campaigns.

\begin{figure*}[tbh!] %
\begin{center}
\includegraphics[width=16.5cm]{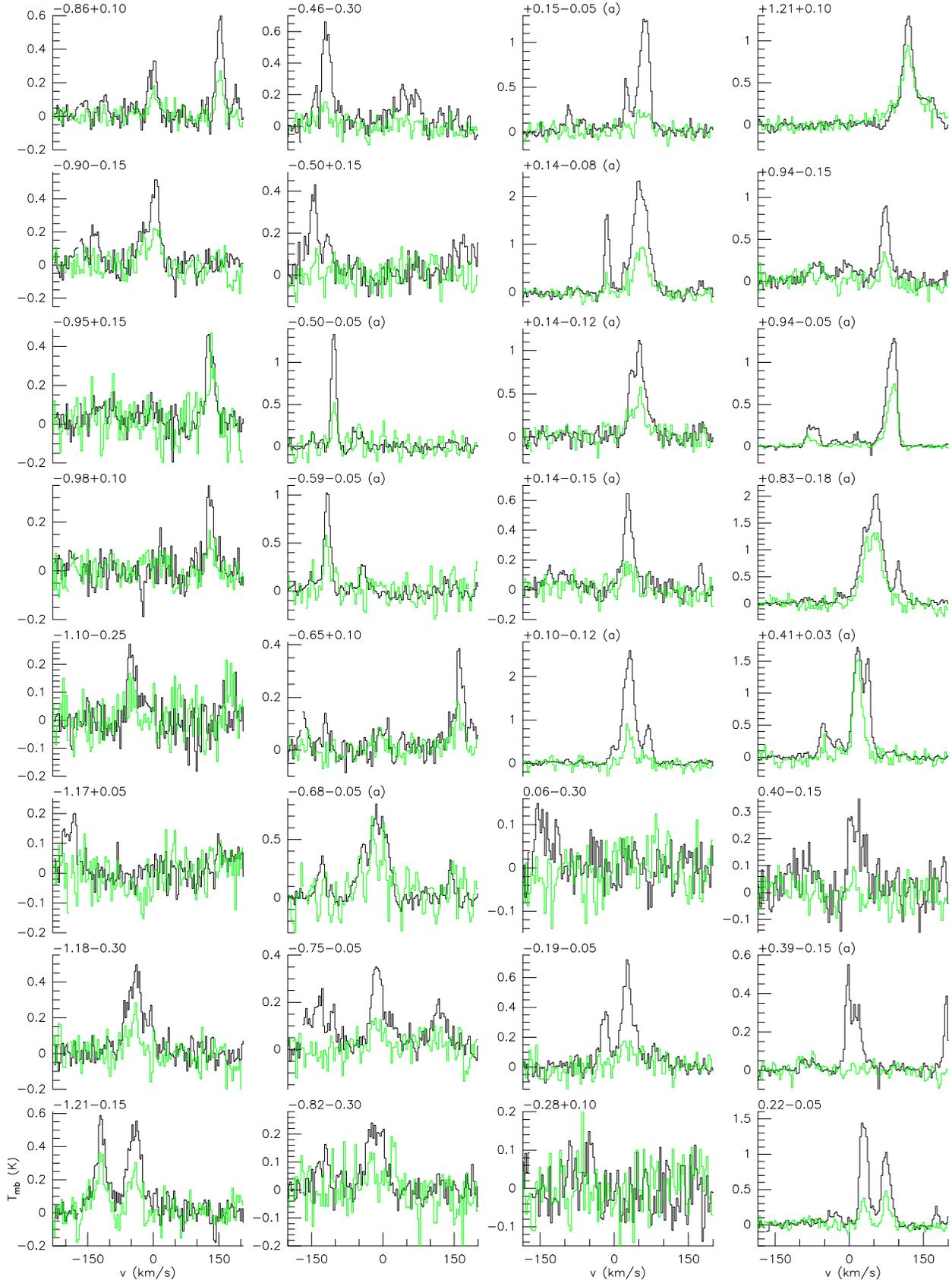}
\caption{CS(2-1) spectra (black lines) and SiO(2-1)  spectra (multiplied by 2, green lines) towards  all the observed  positions, which are  written in the upper-left corner (in Galactic coordinates).
The positions observed with the 30m telescope are indicated with a label (a).
The velocity scale is the same for all the spectra in the same column.}
\label{fig_spectra}
\end{center}
\end{figure*}
\addtocounter{figure}{-1}
\begin{figure*}[tbh!] %
\begin{center}
\includegraphics[width=16.5cm]{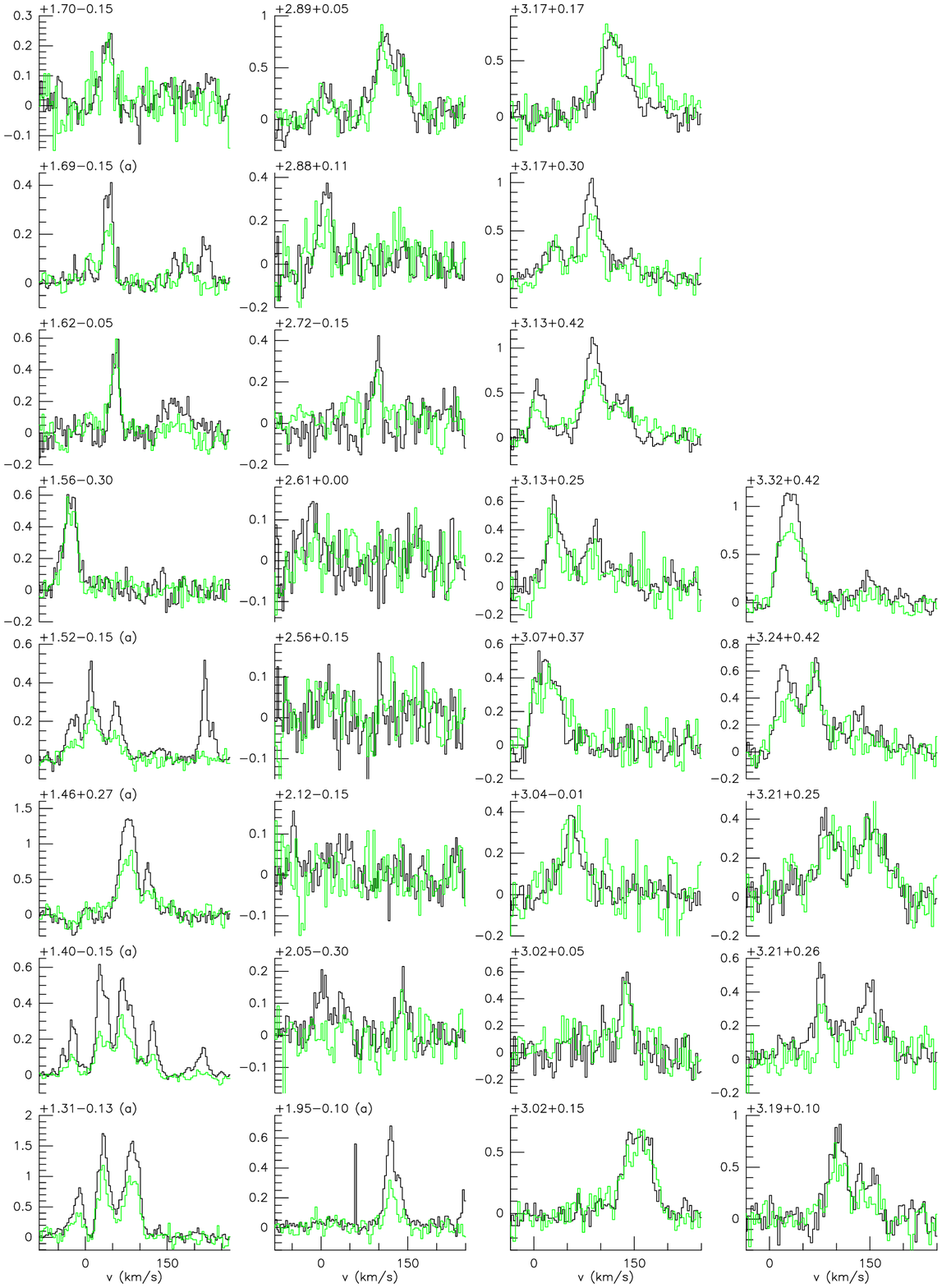}
\caption{Continuation.}
\label{fig_spectra}
\end{center}
\end{figure*}

\begin{figure*}[tbhp!] %
\begin{center}
\caption{In grey-levels and contours we show the longitude-velocity diagram obtained with the CO(1-0)
data by \cite{Bally87} at different latitude ranges. We also indicate the different kinematical features discussed in this paper (note that not all the features that are seen in different 
panels are labeled in all of them). Blue triangles and  bars represent the peak velocity and the full width at half maximum (FWHM)  of the sources detected in CS. Black points and  bars show the peak velocity and FWHM of the sources that have been detected in CS and SiO.}
\label{fig_bally}
\end{center}
\end{figure*}

\section{Results}

Figure \ref{fig_spectra} shows the CS and SiO spectra obtained towards all the observed positions and  Table \ref{tab_results} gives the  velocity, widths and intensities of the CS lines as derived from Gaussian fits  to the 161 lines detected. 
 When the SiO lines are detected, the velocities and the linewidths are in agreement within errors with those of the CS lines and Table \ref{tab_results} shows the SiO to CS intensity ratio
without any correction by the slightly different telescope beams. For the clouds where the SiO line has not been detected, Table \ref{tab_results} gives upper limits to the SiO(2-1)/CS(2-1) line ratio as derived from the 3$\sigma$ limits to the SiO intensity. The results are also displayed graphically in Fig. \ref{fig_bally}. Blue triangles and  bars represent the peak velocity and the full width at half maximum (FWHM)  of the sources detected in CS. Black points and  bars show the peak velocity and FWHM of the  sources  detected in CS and SiO.

We have detected CS emission in  all the kinematical structures of the GC $lv$-diagram (Fig \ref{fig_bally}, Table \ref{tab_results}). The large linewidths measured (8-64 \kms)  are  typical  of Galactic center clouds. The CS intensities range between the high values (0.5-1.5 K) measured for the sources in the $l=1.3\deg$-complex, the Clump 2 and the GCR clouds to 0.1-0.3 K in the  faint structures  that constitute the upper and lower edges of the $lv$-diagram (K, M, R), including  the vertical appendix of the component  K at $l \sim 1.7\deg$, which can be the signature of mass transfer from the elongated orbits to the GCR \citep{Fux99}. The clouds in the right edge of the $lv$-diagram  (L, N, P) and those in the structure J exhibit intermediate CS intensities (0.2-0.6 K). The observations presented here  demonstrate that there are dense CS-emitting clouds  in all the kinematical components seen in the  CO $lv$-diagram and not only in the GCR and the $l=1.3\deg$-complex clouds detected by \cite{Bally87}.

We have also detected SiO(2-1) emission in at least a few clouds of all the kinematical components.  The SiO(2-1) line has been detected  in all the observed clouds of the $l=1.3\deg$-complex and Clump 2.  The SiO/CS line ratio is in the range 0.3-0.6.  The SiO line has also been detected in two clouds of the component K: one cloud near the CO maximum at $l   \sim 1.4$ and one cloud  in the vertical appendix at $l   \sim 1.7\deg$. The SiO/CS ratio is 0.14 and 0.5, respectively. SiO emission has also been detected in the lower edge (component M) and the negative longitude extremity of the $lv$-diagram  (components P and N, including a few clouds with different velocities along the structure N). The SiO/CS ratios are in the range 0.2-0.5.  Similar line ratios have been measured towards several positions in the structures J and L.  In most of the clouds where the SiO line has not been detected the CS is indeed rather weak ($\sim 0.1$ K) and the upper limits to the SiO/CS line ratio are not significant.

\begin{table*}[tbh!]
\caption{Summary of the $^{13}$CO (2-1) and (1-0) observations.
  Intensity of the   $^{13}$CO (1-0) line ($T_{mb}$) and $^{13}$CO (2-1)/(1-0) intensity ratio
calculated from the $T_{mb}$ temperatures without any
correction for the different beam sizes. When the  $^{13}$CO (2-1) line has not been detected we have used  3$\sigma$ upper limits to  calculate the 2-1/1-0 ratio. The LVG results for $T_K$=20 K are also displayed. 
For the sources where the $^{13}$CO (2-1)/(1-0) ratio could not be determined, we have assumed a H$_ 2$ density of  $n=10^{3-4.5}$ \cmmt. 
The SiO column density,  $N$(SiO), has been calculated for a H$_2$ density of $10^{4.5}$ \cmmt .
The SiO abundance relative to H$_2$, $X$(SiO), has been calculated using a $^{13}$CO abundance
of  $5\,10^{-6}$  (see text).}
\label{tab_lvg} \footnotesize
\begin{center}
\begin{tabular}{lllllllllll}
\hline
 $l $  & $b$   & $v$   & Feature  &     $^{13}$CO (1-0)     &   2-1/1-0  & log($n$(H2)) & $N(^{13}$CO)   & $N$(SiO)          & $X$(SiO)\\
 deg & deg & km/s &        &    $T_{mb}$(K)          &             & log(\cmmt) &$10^{16}$ \cmmd & $10^{13}$ \cmmd &  $10^{-9}$   \\
\hline
$ -0.5030$ &$ -0.0500$ &  144.5 & L   & 0.8$\pm$0.2 & 0.7$\pm$0.4 & ${3.0-4.0}$    & 1.1     & $<1.1$      & $<5$  \\
 $-0.5900$ &$ -0.0500$ &  152.8 & L   & 0.5$\pm$0.2 & $<$1.7      & $<{4.3}$       & 0.38    & $<0.8$      & $<9.9$ \\
$ -0.6880$ &$ -0.0500 $&  142.5 & L   & 1.0$\pm$0.7 & 0.7$\pm$0.5 & ${3.-3.5}$     & 0.6     & 1.0         & 7.9  \\
  0.9452 & $-0.0500 $&   86.2 & L   & 4.5$\pm$0.3 & 1.0$\pm$0.1 & ${3.4-3.6}$    & 12      & 7.3         & 3.0 \\
  0.1500 &$ -0.0500$ &  154.7 & K   & 1.2$\pm$0.1 & 0.6$\pm$0.1 & ${3-3.5}$      & 0.45    & $<0.3$      & $<3.1$ \\
  1.3100 & $-0.1300$ &  204.3 & K   & 1.1$\pm$0.3 &  ...        & ...            & $2-4$     & $<0.2$      & $<0.33$  \\
  0.3899 & $-0.1500$ &  195.8 & K   & 0.8$\pm$0.3 & ...         & ...            & 0.8     & $<2.5$      & $<15$\\
  1.4027 & $-0.1500$ &  214.3 & K   & 1.0$\pm$0.1 & 0.6$\pm$0.1 & ${3.-3.4}$     &  2      & 0.4         & 0.88\\
  1.5220 &$ -0.1500 $&  219.5 & K   & 1.6$\pm$0.3 &  ...        & ...            & 2.5     & 0.2         & 0.4 \\
  0.3899 & $-0.1500$ &  $-92.2$ & M   & 0.5$\pm$0.3 & 1.3$\pm$0.8 & ${3.5-4.3}$    &  $0.2-1$  & 0.3         & $1.3-6.8$\\
  0.3899 & $-0.1500 $&  $-71.5$ & M   & 0.3$\pm$0.2 & $<$2.3      & $<{4.3}$       & $>0.6$  & 0.6         & $<5$\\
  0.9452 &$ -0.0500$ &  $-81.0$ & M   & 1.0$\pm$0.3 & 1.0$\pm$0.4 & ${3.-3.8}$     & 1.1     & 0.7         & 3.2\\
  0.9452 & $-0.0500$ &  $-65.9$ & M   & 1.6$\pm$0.3 & 0.7$\pm$0.2 & ${3.-3.5}$     & 2.8     & 0.4         & 0.63\\
$ -0.6880$ & $-0.0500$ &  $-41.7$ & P   & 0.9$\pm$0.4 & 1.5$\pm$0.7 & $3.5-5 $       & 1.7     & 2.7         & 7.9 \\
 $-0.6880$ & $-0.0500$ &  $-17.5$ & P   & 1.0$\pm$0.4 & 2.2$\pm$0.9 & ${3.5-5}$      & 1.0     & 3.2         & 16 \\
 $-0.6880$ & $-0.0500$ &    2.3 & P   & 2.0$\pm$0.4 & 1.5$\pm$0.3 & ${3.5-5}$      & 1.0     & 5.0         & 25 \\
 $-0.5900$ & $-0.0500$ &  $-41.3$ & P   & 1.6$\pm$0.2 & 0.9$\pm$0.2 & ${3.5-4}$      & 2.2     & $<1.4$      & $<3.2$  \\
  1.3100 & $-0.1300$ &  $-13.8$ & 1.3 & 2.0$\pm$0.3 & 0.8$\pm$0.2 & ${3.5-4.3}$    & 5.7     & 0.2         & 0.16 \\
  1.3100 & $-0.1300$ &   33.7 & 1.3 & 1.9$\pm$0.3 & 1.0$\pm$0.2 & ${3.5-4.3}$    & 5.7     & 9.5         & 8.3  \\
  1.3100 & $-0.1300$ &   85.0 & 1.3 & 2.4$\pm$0.3 & 1.1$\pm$0.2 & ${3.5-4.3}$    & 7.5     & 5.3         & 3.5  \\
  1.3100 & $-0.1300$ &  135.0 & 1.3 & 0.5$\pm$0.3 & $<$1.7      & $<10^4$        & 0.4     & $<0.1$      & $<1.3$  \\
  1.4027 & $-0.1500$ &  $-41.8$ & 1.3 & 1.1$\pm$0.1 & 0.6$\pm$0.1 & ${3.}$         & 0.6     & $<0.1$      & $<1.0$ \\
  1.4027 & $-0.1500$ &  $-22.9$ & 1.3 & 1.5$\pm$0.3 & 0.7$\pm$0.2 & ${3-3.5}$      & 1.9     & 0.5         & 1.2\\
  1.4027 & $-0.1500$ &   29.7 & 1.3 & 1.5$\pm$0.2 & 0.9$\pm$0.2 & ${3.3-3.5}$    & $2.2-3.5$ & 1.4         & 2.5\\
  1.4027 & $-0.1500 $&   72.3 & 1.3 & 1.6$\pm$0.2 & 0.9$\pm$0.1 & ${3.3-3.5}$    & 6.6     & 2.3         & 1.7 \\
  1.4027 & $-0.1500$ &  123.7 & 1.3 & 0.6$\pm$0.0 & 1.3$\pm$0.1 & ${3.5-4.0}$    & 0.85    & 0.5         & 3\\
  1.5220 & $-0.1500$ &  $-23.4$ & GCR & 1.1$\pm$0.8 & 0.8$\pm$0.6 & $3.2-4$        & 2.3     & 0.5         & 1.0  \\
  1.5220 & $-0.1500$ &   56.0 & GCR & 1.1$\pm$0.8 & 0.9$\pm$0.7 & $<4.3$         & 3.4     & 0.9         & 1.3  \\
 $-0.6880$ & $-0.0500$ & $-129.1$ & GCR & 2.6$\pm$0.4 & 1.2$\pm$0.2 & $3.5-4.5$      & 4.5     & 1.1         & 1.2 \\
 $-0.5900$ & $-0.0500$ &  132.9 & GCR & 0.3$\pm$0.0 & $<$3.3      & ...            & 16      & $<0.8$      & $<0.25$  \\
 $-0.5900$ & $-0.0500$ & $-148.5$ & GCR & 1.1$\pm$0.2 & 1.0$\pm$0.3 & $3.5-4.5$      & 2.5     & $<2.6$      & $<5.2$ \\
 $-0.5900$ & $-0.0500$ & $-116.7$ & GCR & 5.3$\pm$0.2 & 1.0$\pm$0.1 & $>3.5$         & 9.5     & 3.0         & 1.6 \\
 $-0.5030$ & $-0.0500$ & $-147.4$ & GCR & 1.1$\pm$0.2 & 0.6$\pm$0.3 & $2.8-3.5$      & 0.5     & $0.1-0.8$     & $1.3-8$ \\
 $-0.5030$ & $-0.0500$ &  $-53.5$ & GCR & 1.0$\pm$0.2 & 1.3$\pm$0.4 & $3.5-4.5$      & 1.4     & $0.7-2.1$     & $2.5-7.5$ \\
  0.1500 & $-0.0500$ &   25.5 & GCR & 4.0$\pm$0.4 & 1.2$\pm$0.2 & $>3.5$         & 6       & $<1$        & 0.79 \\
  0.1500 & $-0.0500$ &   61.1 & GCR & 5.3$\pm$0.4 & 1.4$\pm$0.1 & $>3.5$         & 14      & 1.5         & 0.54 \\
  0.8300 & $-0.1800$ &   31.3 & GCR & 5.0$\pm$0.5 & 1.0$\pm$0.1 & $>3.5$         & 16      & 6.4         & 2  \\
  0.8300 & $-0.1800$ &   56.9 & GCR & 0.5$\pm$0.5 & 1.3$\pm$1.5 & $3.5-4.5$      & 1.1     & 8.5         & 30  \\
  0.8300 & $-0.1800$ &   99.1 & GCR & 1.8$\pm$0.5 & 1.0$\pm$0.3 & $3.2-4.5$      & 1.4     & 1.4         & 5  \\
  0.4100 &  0.0300 &  $-51.0$ & GCR   & 1.1$\pm$0.5 & 1.2$\pm$0.6 & ${3.5-4.2}$    & 1.9     & 2.5         & 6.6\\
  0.4100 &  0.0300 &  $-21.9$ & GCR  & 1.5$\pm$0.5 & 0.8$\pm$0.3 & ${3.-3.5}$     & 3.3     & 1.1         & 1.7 \\
  0.4100 &  0.0300 &   17.0 & GCR & 3.9$\pm$0.5 & 1.0$\pm$0.2 & $3.5-4.3$      & 6.8     & 12          & 8.8 \\
  0.4100 &  0.0300 &   37.9 & GCR & 5.1$\pm$0.5 & 1.1$\pm$0.1 & $>3.5$         & 8.2     & $<1.3$      & $< 0.79$\\
  0.1000 & $-0.1200$ &   $-2.0$ & GCR & 0.8$\pm$0.0 & 2.2$\pm$0.7 & $>3.5$         & 1.1     & 0.9         & 4\\
  0.1000 & $-0.1200$ &   30.1 & GCR & 6.0$\pm$0.6 & 1.0$\pm$0.2 &  $3.5-4.3$     & 25      & 8.2         & 1.64\\
  0.1000 & $-0.1200$ &   68.1 & GCR & 1.9$\pm$0.6 & 1.4$\pm$0.6 & $>3.5$         & 3.8     & 0.9         & 1.2 \\
  0.3899 & $-0.1500$ &  $ -3.5$ & GCR & 1.8$\pm$0.2 & 1.5$\pm$0.2 & $>3.7$         & 1.9     & $<0.3$      & $<0.7$\\
\hline
\end{tabular}
\end{center}
\end{table*}

\section{Gas density and cloud stability}

To get further insight in the physical conditions   in the different kinematical structures we have obtained $^{13}$CO (1-0) and (2-1) data for some sources. Table \ref{tab_lvg} shows $^{13}$CO(1-0) line intensities and $^{13}$CO (2-1)/(1-0) line ratios without any correction by the different beams. The line ratio is 0.6-1.5 for most of the sources (it  would be a factor of two higher if the emission were extended and homogeneous). We have used a Large Velocity Gradient (LVG) code to determine the H$_2$ density and the $^{13}$CO column density for two values of the kinetic temperature ($T_K$): 20 and 150 K. Table \ref{tab_lvg} shows the results obtained for a kinetic temperature of 20 K. The measured line ratios imply an H$_2$ densitiy of $10^{3.5}-10^{4.5}$ \cmmt ~ in the GCR clouds and $10^3-10^{3.5}$ \cmmt ~ in the components K or M. This is in perfect agreement with the results of  \cite{Sawada01} and \cite{Martin04}. 
Assuming a kinetic temperature of 150 K \citep{Huttemeister93,Rodriguez00,Rodriguez01} the H$_2$ density would be  a factor  of 3 lower.

In addition, the detection of the CS(2-1) and  SiO(2-1) lines   in all the velocity components  imply  the presence of high densities  \citep[$\geq10^{4}$ \cmmt,][]{Linke80} in the core of the clouds.
Therefore, all the structures seen in the CO $lv$-diagram contain dense molecular gas.  In particular, this is true for  the gas moving in elongated orbits related to the dust lanes (like the components J and K) or about to enter the dust lanes shocks \citep[Clump 2, see also][]{Stark86}.  This result is in agreement with what is found in  external galaxies, where the density contrast between the dust lanes and the nuclear ring is low \citep[see for instance the results on  NGC 5383 by][]{Sheth00}.  

The density of the clouds is directly related to the problem of cloud stability, which is of  crucial interest to understand the star formation process.  Due to the strong tidal forces  in the Galactic center, the clouds must be denser than in the disk of the Galaxy.
 This question has been studied in number of works  \citep{Stark86, Gusten89, Stark89, Stark04},
 For exemple,  \citet{Stark04} have found  that  the minimum density (in \cmmt) for a cloud to withstand the tides can be expressed as:
\begin{equation}
n\approx 10^{3.5} \mathrm{cm}^{-3} \left[  \frac{\kappa}{1000 \, \mathrm{Gyr}^{-1}}  \right] ^2
\end{equation} 
where $\kappa$ is the epicyclic frequency, which can be estimated to 1036 Gyr$^{-1}$ for a radius of 150 pc and to 506 Gyr$^{-1}$ for a radius of 450 pc \citep{Stark04}.
Assuming a sun-GC  distance of 8.5 kpc, 150 pc  is  the radius of the GCR.
Therefore,  the critical density is $\sim3\,10^3$ \cmmt.
The clouds with non-circular velocities are located at larger galactocentric radius than the GCR, therefore the critical density is  lower. For instance, the critical density for a cloud located at 450 pc is a factor of $\sim 4$ lower than for a cloud located at 150 pc \citep{Stark04}.
Therefore,  the CS and SiO clouds in non-circular orbits are gravitationally bound and 
 the dust lanes do not contain diffuse gas only \citep[see also][]{Stark04}.

\begin{figure}[tbhp!] %
\begin{center}
\includegraphics[angle=-90,width=8.5cm]{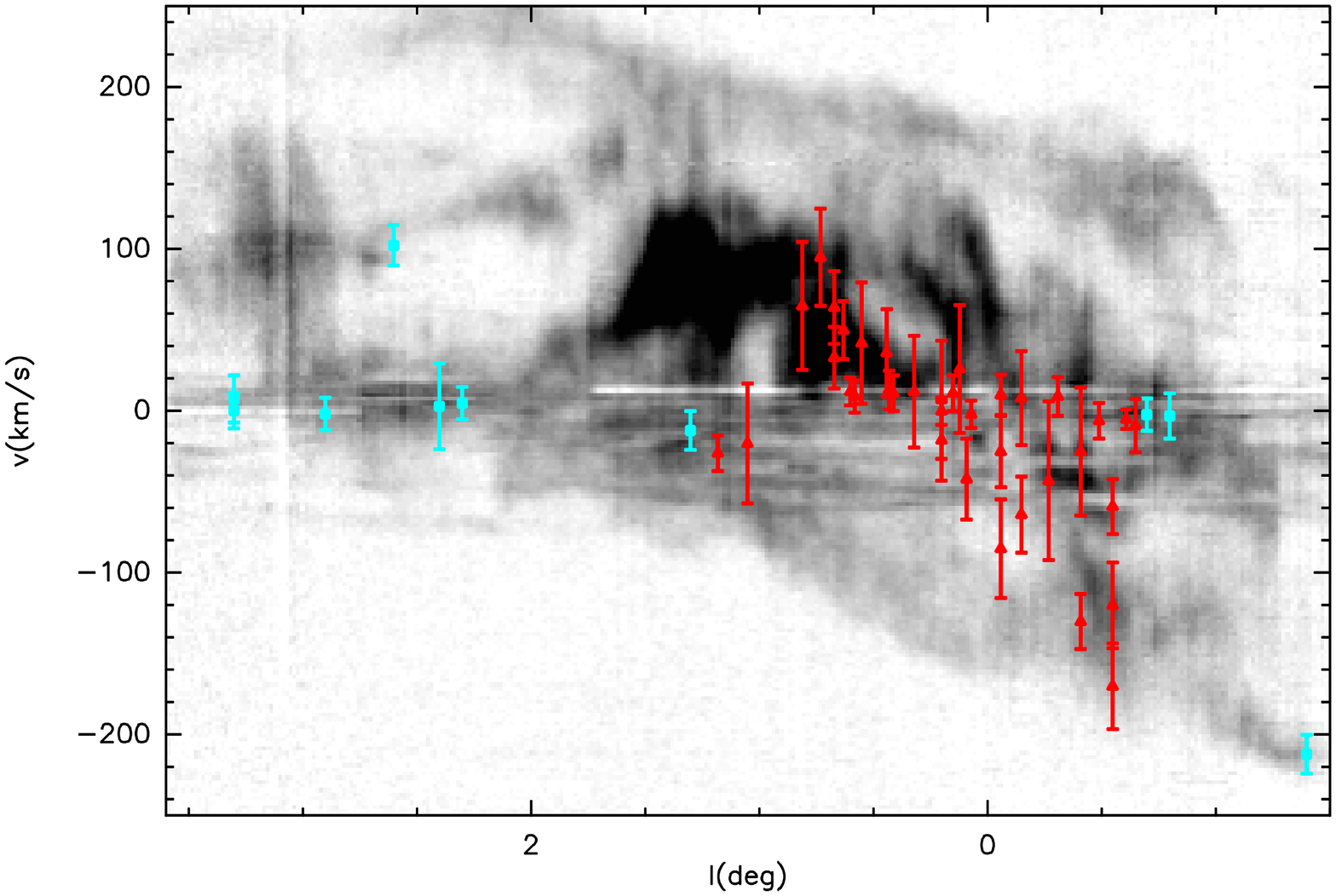}
\caption{ In grey-levels  we show the longitude-velocity diagram obtained integrating the CO(1-0)
data by \cite{Bally87}  in the latitude range $-0.3^\circ <b<0.3^\circ$.  Red triangles and  bars represent the peak velocity and the full width at half maximum (FWHM)  of the radio recombination lines detected in the longitude range $1.18^\circ >l>-0.55^\circ$ \citep{Pauls75}.  Blue squares  and  bars represent the peak velocity and the FWHM  of the radio recombination lines detected at $l>1.18^\circ$ and $l<-0.55^\circ$ \citep{Paladini03}.}
\label{fig_hii}
\end{center}
\end{figure}

 However, there are no signs of star formation in the dust lanes. 
The surveys of hydrogen recombination lines (at centimeter wavelengths) in the inner 2 degrees \citep{Pauls75} show that the H{\sc ii} regions  are mostly concentrated  along a line that runs through the origin of the $lv$ plane indicating that the star formation is confined to the GCR (see Fig. \ref{fig_hii}). We have also looked for H{\sc ii} regions at larger longitudes in the catalog by \citet{Paladini03}.
Figure \ref{fig_hii} shows that there are two  H{\sc ii} regions at longitude $l \sim -0.6^\circ$ associated with the structure P (or to the spiral arms in the line of sight) and one ionized region at $l=-1.4^\circ$ associated with Sgr E (and therefore to the GCR).
At  longitudes $l>1.2^\circ$ all the known H{\sc ii} regions are most likely associated with the spiral arms except one source with $l=2.6^\circ$ and $v=102$ \kms\ that could be associated with the structure J. However this structure is basically seen at latitudes $b<-0.1^\circ$  (see Fig. \ref{fig_bally}) and the H{\sc ii} region is located at a latitude $b=0.1^\circ$ \citep{Paladini03}.
Therefore, the data available show no signs of star formation in the cloud complexes associated to the dust lanes (Clump 2, J, K, M). 
However, we can not rule out the presence of ultracompact  H{\sc ii} regions (UC--H{\sc ii}) embedded in the dense clumps since very young H{\sc ii} regions could be optically thick  at centimeter wavelengths.  The presence of isolated UC--H{\sc ii} regions is a possibility that can only be ruled out by deep high resolution searches for recombination line emission at millimeter wavelengths. Nevertheless,  such  very young UC--H{\sc ii} regions are  usually associated with more evolved H{\sc ii} regions that should have been detected by \citet{Pauls75} and the surveys compiled by  \citet{Paladini03}. In summary, with the available recombination line data data there are no signs of massive star formation in the dust lanes.

Indeed,  the lack of star formation along the dust lanes
is  common in the observations of  early-type barred galaxies, where there is  star formation in the  nuclear ring and at the  extremities of the bar, but  there are no signs of star formation in the dust lanes along the bar \citep[see for instance][]{Tubbs82,Reynaud98, Sheth00}.
The inhibition of the star formation can be an effect of the high velocities, relative to the gas in the dust lanes, of the dense clouds whose orbits intersect the dust lanes \citep{Tubbs82}. These velocities are much higher in barred than in normal galaxies. The models by \cite{Tubbs82} showed that the clouds that enter the dust lanes at velocities higher than a critical value ranging between 20 and 60 \kms ~ must be inhibited from forming stars to explain the observations of NGC 5383. The high velocity can cause a quick compression of the clouds followed by a rapid expansion which disperses the majority of the cloud. This mechanism has also been invoked by \cite{Reynaud98}  to explain the observations of NGC 1530.  This mechanism could also inhibit the star formation in the gas with non-circular velocities in the GC. 

\section{SiO abundances and the origin of the SiO emission}

As already mentioned,  the SiO/CS line ratio is expected to depend mainly on the relative abundance of the two molecules. For example, in the cloud sample studied by \cite{Huttemeister93b} and \cite{Huttemeister98}, which is basically composed by clouds of the  GCR and the $l=1.3\deg$-complex, the SiO abundance shows a clear trend to increase from $2 \, 10^{-10}$ to $10^{-8}$ with increasing SiO(2-1)/CS(2-1) line ratios in the range 0.1-0.6. These should also be the SiO abundances of the sources of this paper with detected SiO.

In fact, we have derived the  SiO abundances for some sources  of our sample using the $^{13}$CO data and LVG calculations.  
Table \ref{tab_lvg} shows the SiO column densities obtained with the typical temperature ($T_K=20$~K)  and density ($n(H_2) \sim 10^{4.5}$ \cmmt) of the SiO clouds in the GC \citep[][]{Martin-Pintado97, Huttemeister98}. 
 The abundance of $^{13}$CO relative  to H$_2$ is 5\,10$^{-6}$ in the Galactic center \citep[see][and references therein]{Huttemeister98}. 
Therefore,  the   SiO to $^{13}$CO column density ratios 
 imply abundances of SiO relative to H$_2$ of $(0.5-8)\,10^{-9}$ for the sources with detected SiO, including those in the structures K and M. These abundances are in agreement with  the results of \cite{Huttemeister98} and \cite{Martin-Pintado97, Martin-Pintado00} for clouds located mainly in the GCR and the $l=1.3\deg$-complex. For one source in the Sgr B2 area ($l=0.83^\circ, \, b=-0.18^\circ, \, v=57$ \kms) and two in the feature P we have even measured a SiO  abundance of $\sim 10^{-8}$, which is comparable to the highest abundances derived by \cite{Huttemeister98} in the $l=1.3^\circ$-complex.


The SiO abundance shows extreme variations in the Galaxy. The abundance is $<5\, 10^{-12}$ in dark clouds \citep{Ziurys89} and $10^{-11}-10^{-10}$ in diffuse translucent clouds and in spiral arms clouds \citep{Greaves96, Turner98,Lucas00}. \cite{Schilke01} have measured an abundance of $\sim 10^{-11}$ in the photo-dissociation region (PDR) of the Orion bar and S140. In contrast, the SiO abundance in shocked regions associated to star formation  is $\sim 10^{-8}$ \citep{Martin-Pintado92, Bachiller97}. 
In these regions the Si and SiO abundances in the gas phase are increased by the processing of the dust grains by shock waves \citep{Schilke97}.
Thus, the SiO abundances in the non-circular velocity GC clouds measured in this work and those derived in previous works for other GC clouds \citep{Martin-Pintado97, Martin-Pintado00, Huttemeister98}, are close to the values found in gas shocked  by molecular outflows from protostars.  
Measurements of the SiO abundances in the nuclei of  external galaxies like NGC 253, M82, NGC 1068 or IC 342 \citep{Garcia-Burillo00, Garcia-Burillo01, Usero04, Usero05}  give values  from $10^{-10}$ to a few $10^{-9}$,  comparable to those measured in the GC.

Two main scenarios have been invoked to explain the widespread high abundances of SiO in  the GC and the center of the nearby galaxies cited above: 

(i) Grain sputtering by shocks waves  of different origins  \citep{Martin-Pintado97}. \cite{Huttemeister98} have found evidence of a link between the large scale Galactic dynamics and the origin of the shocks in the GC (high SiO abundances  in  the $l=1.3^\circ$-complex, which  they proposed to be due to the interaction of $x_1$ and $x_2$ orbits). \cite{Garcia-Burillo00} and \cite{Usero05} also use shocks related to the galactic dynamics to explain the SiO emission in NGC 253 and IC 342.

(ii) More recently, a second mechanism has been proposed for regions in the vicinity of X-ray sources: the destruction of small silicate grains by energetic X-rays. This mechanism could play a role in the GCR \citep{Martin-Pintado00} and close to the AGN in NGC 1068 \citep{Usero04}.

As it is well known, the arena of energetic phenomena is in the inner regions of the GC  (birth and death of  massive stars, energetic radiation, shocks,...) making it very complex to interpret the SiO observations of the GCR clouds .   
In contrast, as discussed above, there are no signs of star formation in the dust lanes and the SiO abundances are as high as in the other GC sources.
The high SiO abundances measured  in  the GC clouds with non-circular velocities can be   due to the strong shocks that created the dust lanes.
This scenario has also been proposed by \cite{Meier05} to interpret the detection of HNCO and CH$_3$OH emission along the bar of IC 342.
\cite{Usero05} have  also  detected SiO emission in the dust lane along the bar of IC 342 and  they have measured high SiO abundances  of a few $10^{-9}$  (about a factor of 5-10 higher than in the nuclear ring). They interpret their findings as the effect of large-scale shocks driven by cloud-cloud collisions along the potential well of the bar.


\section{Conclusions}

We have shown the presence of dense gas in all the kinematical components of the Galactic center, including the structures with non-circular  velocities like  the dust lanes along the bar. The gas density of these clouds are higher than the critical density for cloud stability. However, in the GC there are no signs of star formation except in the nuclear ring (the GCR).  
 The high relative velocities and shear expected in other kinematical structures, in particular in those associated with the dust lanes, could inhibit the star formation. 
This is the mechanism  proposed to explain the lack of star formation in the dust lanes along the bar of  Galaxies  like NGC 5383 and NGC 1530.  The high  measured abundances of one of the most representative molecules of the GC shock chemistry (SiO) in  the cloud components associated with the dust  lanes and elongated orbits supports this scenario. Furthermore, it gives us important hints on the origin of the emission of this molecule. The origin  of the shocks and therefore the GC molecular chemistry can indeed be strongly influenced by the global dynamics of the Galaxy. More observations of the molecular chemistry in face-on barred galaxies would be very interesting for a better understanding of the links between the gas dynamics and the molecular chemistry.

\begin{acknowledgements}
We thank Jonathan Braine for a critical reading of the manuscript and Sergio Martin for helping with the 30m observations.
NJR-F has been partially supported by a Marie Curie  Fellowship of
the European Community program ``Improving Human Research
Potential and the Socio-economic Knowledge base''  under contract
number HPMF-CT-2002-01677.  
JM-P Acknowledges support by the Spanish Ministerio de Educaci\'on y Ciencia 
under projects AYA 2002-10113-E, AYA~2003-02785-E, and ESP~2004-00665.
\end{acknowledgements}





\begin{table}[p]
\caption{Galactic longitude and latitude (in degrees),  velocity (km\,s$^{-1}$), widths (km\,s$^{-1}$), and intensities (K) of the CS lines. In addition, the SiO to CS intensity ratio is also shown. The numbers in  parentheses are the uncertainties on the last digit. The positions observed with the IRAM 30m telescope are indicated with a 
superscript. All the other positions have been observed with the 12m telescope.}
\label{tab_results}
\begin{center}
\begin{tabular}{llllll}
\hline
$l$ & $b$ & $v$  & $\Delta v$  & I(CS) & $\frac{I(SiO)}{I(CS)}$ \\
\hline
\multicolumn{6}{c}{\bf l=1.3-complex region} \\
+1.403$^{a}$ & -0.150 &   29(1) & 22(1) & 0.58(3) &     0.19(2)   \\
+1.310$^{a}$ & -0.130 &   33(1) & 23(1) & 1.58(9) &     0.33(3)   \\
+1.403$^{a}$ & -0.150 &  123(2) & 16(1) & 0.36(3) &     0.17(3)   \\
+1.403$^{a}$ & -0.150 &   72(1) & 33(1) & 0.49(3) &     0.33(3)   \\
+1.310$^{a}$ & -0.130 &   85(1) & 32(1) & 1.53(9) &     0.33(3)   \\
+1.310$^{a}$ & -0.130 &  135(1) & 11(4) & 0.16(9) & \le 0.42      \\
+1.218       & +0.100 &  119(4) & 56(7) & 0.39(3) &     0.63(8)   \\
+1.218       & +0.100 &  119(2) & 17(1) & 0.91(3) &     0.24(3)   \\
+1.218       & +0.100 &  163(3) & 28(4) & 0.26(3) &     0.34(11)  \\
+1.460$^{a}$ & +0.270 &   78(1) & 32(5) & 1.43(7) &     0.28(2)   \\
+1.460$^{a}$ & +0.270 &  116(2) & 16(1) & 0.64(7) &     0.25(6)   \\
+1.698$^{a}$ & -0.150 &    3(3) & 15(3) & 0.09(3) &     0.7(3)    \\
+1.522$^{a}$ & -0.150 &   11(2) & 24(1) & 0.41(4) &     0.27(4)   \\
 +1.560       & -0.300 &  -26(1) & 29(1) & 0.54(4) &     0.46(6) \\
 +1.522$^{a}$ & -0.150 &  -23(2) & 23(1) & 0.22(4) &     0.18(6) \\
 +1.403$^{a}$ & -0.150 &  -41(1) &  8(1) & 0.15(3) & \le 0.11 \\
 +1.403$^{a}$ & -0.150 &  -22(2) & 15(1) & 0.30(3) &     0.20(4) \\
 +1.310$^{a}$ & -0.130 &  -13(2) & 23(1) & 0.73(9) &     0.25(7) \\
+1.620       & -0.050 &  152(5) & 30(8) & 0.10(6) &     0.6(6)    \\
+1.708       & -0.150 &   40(2) & 25(2) & 0.20(2) &     0.57(9)   \\
+1.698$^{a}$ & -0.150 &   42(3) & 19(3) & 0.39(3) &     0.31(4)   \\
+1.522$^{a}$ & -0.150 &   56(2) & 26(1) & 0.27(4) &     0.22(5)   \\
+1.620       & -0.050 &   55(5) & 16(1) & 0.54(5) &     0.57(7)   \\
\hline
\multicolumn{6}{c}{\bf Clump 2}       \\
+3.329       & +0.425 &   30(1) & 33(1) & 1.22(6) &     0.34(3)   \\
+3.329       & +0.425 &  150(1) & 32(4) & 0.24(6) &     0.3(2)    \\
+3.235       & +0.425 &   68(3) & 22(3) & 0.59(6) &     0.47(7)   \\
+3.235       & +0.425 &   26(3) & 31(3) & 0.63(6) &     0.32(5)   \\
+3.219       & +0.250 &   84(1) & 34(3) & 0.38(6) &     0.30(10)  \\
+3.219       & +0.250 &  152(1) & 58(4) & 0.33(6) &     0.47(13)  \\
+3.210       & +0.263 &   77(1) & 27(2) & 0.45(6) &     0.32(8)   \\
+3.210       & +0.263 &  146(1) & 55(3) & 0.33(6) &     0.29(11)  \\
+3.193       & +0.100 &  101(1) & 27(2) & 0.78(6) &     0.35(7)   \\
+3.193       & +0.100 &  142(1) & 27(3) & 0.46(6) &     0.22(11)  \\
+3.172       & +0.175 &  119(1) & 45(1) & 0.72(5) &     0.44(7)   \\
+3.172       & +0.175 &  183(2) & 31(6) & 0.16(5) &     0.9(4)    \\
+3.171       & +0.300 &   29(1) & 29(2) & 0.33(5) &     0.41(12)  \\
+3.171       & +0.300 &   84(2) & 33(1) & 0.92(5) &     0.32(4)   \\
+3.171       & +0.300 &  143(1) & 37(4) & 0.19(5) &     0.3(2)    \\
+3.138       & +0.425 &    7(1) & 23(2) & 0.54(4) &     0.32(5)   \\
+3.138       & +0.425 &   88(2) & 28(1) & 1.03(4) &     0.34(3)   \\
+3.138       & +0.425 &  130(1) & 30(2) & 0.46(4) &     0.27(6)   \\
+3.136       & +0.250 &  131(4) & 32(7) & 0.11(5) &     0.4(4)    \\
+3.136       & +0.250 &   88(1) & 30(3) & 0.35(5) &     0.29(12)  \\
+3.136       & +0.250 &   31(2) & 29(1) & 0.53(5) &     0.47(9)   \\
+3.074       & +0.375 &   18(2) & 37(1) & 0.52(6) &     0.38(8)   \\
+3.047       & -0.013 &   59(1) & 34(2) & 0.35(5) &     0.4(2)    \\
+3.020       & +0.150 &  156(1) & 46(1) & 0.67(6) &     0.45(7)   \\
+3.020       & +0.050 &  136(1) & 17(1) & 0.62(4) &     0.36(6)   \\
+2.895       & +0.050 &  116(1) & 56(3) & 0.74(2) &     0.39(2)   \\
\hline
\end{tabular}
\end{center}
\end{table}

\addtocounter{table}{-1}
\begin{table}[p]
\caption{Continuation.}
\begin{tabular}{llllll}
\hline
$l$ & $b$ &  $v$  & $\Delta v$  & I(CS) & $\frac{I(SiO)}{I(CS)}$\\
\hline
\multicolumn{6}{c}{\bf Galactic center ring} \\
\multicolumn{3}{l}{Sgr E} \\
 -1.178       & +0.050 & -182(2) & 21(8) & 0.15(2) & \le 0.87 \\
 -1.178       & +0.050 & -203(2) & 14(3) & 0.16(2) & \le 0.87 \\
\multicolumn{3}{l}{ Sgr B} \\
 +0.830$^{a}$ & -0.180 &   31(2) & 32(3) & 1.12(7) &     0.29(5) \\
 +0.830$^{a}$ & -0.180 &   56(1) & 27(1) & 1.71(7) &     0.32(3) \\
\multicolumn{4}{l}{ Sgr A } \\
 +0.227       & -0.050 &   74(2) & 21(1) & 0.97(5) &     0.21(2) \\
 +0.150$^{a}$ & -0.050 &   25(2) & 13(1) & 0.52(6) & \le 0.19    \\
 +0.150$^{a}$ & -0.050 &   61(2) & 23(2) & 1.30(6) &     0.08(1) \\
 +0.145$^{a}$ & -0.080 &   52(1) & 36(2) & 2.21(9) &     0.20(2) \\
 +0.145$^{a}$ & -0.150 &   28(1) & 22(1) & 0.59(3) &     0.12(4) \\
 +0.145$^{a}$ & -0.120 &   32(2) &  8(1) & 0.44(7) &     0.30(7) \\
 +0.145$^{a}$ & -0.120 &   50(1) & 29(1) & 0.97(7) &     0.25(3) \\
 +0.100$^{a}$ & -0.120 &   30(3) & 25(3) & 2.53(7) &     0.15(1) \\
 +0.100$^{a}$ & -0.120 &   68(3) & 15(3) & 0.85(7) & \le 0.16    \\
\multicolumn{4}{l}{ Other sources} \\
+0.830$^{a}$ & -0.180 &   99(1) & 17(1) & 0.67(7) & \le 0.20      \\
+0.945             & -0.150 &   73(2) & 17(1) & 1.18(6) &     0.21(2)   \\
+0.945$^{a}$ & -0.050 &   86(2) & 23(2) & 1.31(8) &     0.28(2)   \\
+0.410$^{a}$ & +0.030 &   17(3) & 20(3) & 1.84(4) &     0.41(2)\\
+0.410$^{a}$ & +0.030 &   37(3) & 13(3) & 1.38(4) & \le 0.10\\
 +0.410$^{a}$ & +0.030 &  -51(3) & 19(3) & 0.52(4) &     0.35(7)\\
 +0.410$^{a}$ & +0.030 &  -21(3) & 21(3) & 0.26(4) &     0.3(2)\\
+0.401       & -0.150 &   12(1) & 35(3) & 0.24(4) & \le 0.22\\
+0.390$^{a}$ & -0.150 &   -3(1) &  9(2) & 0.42(2) & \le 0.11\\
+0.390$^{a}$ & -0.150 &   13(1) & 28(4) & 0.30(2) & \le 0.15\\
+0.227       & -0.050 &   29(1) & 18(2) & 1.51(5) &     0.10(1) \\
+0.227       & -0.050 &  -30(2) & 28(4) & 0.14(5) & \le 0.51\\
 
+0.150$^{a}$ & -0.050 &  -92(2) & 13(2) & 0.27(6) & \le 0.33\\
+0.150$^{a}$ & -0.050 &  -65(1) & 14(3) & 0.14(6) & \le 0.54\\
+0.145$^{a}$ & -0.080 &  -14(1) & 10(1) & 1.68(9) & \le 0.08\\ 
+0.100$^{a}$ & -0.120 &   -2(3) & 14(3) & 0.35(7) & \le 0.35\\
 -0.198       & -0.050 &   27(2) & 27(1) & 0.63(5) &     0.11(3)\\
 -0.198       & -0.050 &   85(3) & 33(8) & 0.10(5) &     0.5(3)\\
 -0.198       & -0.050 &  -18(1) & 21(2) & 0.35(5) & \le 0.17\\
 -0.460       & -0.300 &   45(4) & 34(9) & 0.18(4) & \le 0.25\\
 -0.460       & -0.300 &   72(2) & 18(4) & 0.18(4) & \le 0.25 \\
 -0.503$^{a}$ & -0.050 &  -53(1) & 21(2) & 0.16(4) &     0.9(5)    \\
-0.503$^{a}$ & -0.050 & -147(1) &  8(3) & 0.17(4) &     0.6 (5) \\
 -0.503$^{a}$ & -0.050 & -103(1) & 11(2) & 1.42(4) &     0.18(5) \\
 -0.590$^{a}$ & -0.050 & -148(2) & 25(6) & 0.12(2) & \le 1.24    \\
 -0.590$^{a}$ & -0.050 & -116(2) & 14(1) & 1.02(2) &     0.28(1) \\
 -0.688$^{a}$ & -0.050 & -129(1) & 17(2) & 0.31(4) & \le 0.61 \\
 -0.754       & -0.050 & -132(2) & 61(4) & 0.17(4) & \le 0.25 \\
\multicolumn{3}{l}{ F} \\
 -0.655       & +0.100 & -165(2) & 20(4) & 0.14(3) & \le 0.24 \\
 -0.655       & +0.100 & -117(2) & 15(4) & 0.10(3) & \le 0.33 \\
 -0.503$^{a}$ & +0.150 & -145(6) & 21(2) & 0.59(5) & \le 0.17    \\
 -0.503$^{a}$ & +0.150 & -112(1) & 20(3) & 0.28(5) & \le 0.33    \\
 -0.285       & +0.100 &  -87(4) & 20(9) & 0.07(3) & \le 0.77\\
 -0.285       & +0.100 &  -51(4) & 22(7) & 0.09(2) & \le 0.72\\
\hline
\end{tabular}
\end{table}

\addtocounter{table}{-1}
\begin{table}[p]
\caption{Continuation.}
\begin{center}
\begin{tabular}{llllll}
\hline
$l$ & $b$ & $v$  & $\Delta v$  & I(CS) & $\frac{I(SiO)}{I(CS)}$ \\
\hline
  \multicolumn{6}{c}{\bf L } \\
 -0.503$^{a}$ & -0.050 &  144(3) & 17(4) & 0.07(4) & \le 1.26 \\
 -0.590$^{a}$ & -0.050 &  132(1) &  8(1) & 0.12(2) & \le 1.24 \\
 -0.590$^{a}$ & -0.050 &  152(1) &  7(2) & 0.10(2) & \le 1.44 \\
 -0.688$^{a}$ & -0.050 &  142(1) & 10(1) & 0.30(4) & \le 0.63 \\
 -0.754       & -0.050 &  118(2) & 41(5) & 0.16(4) & \le 0.26 \\
 -0.503$^{a}$ & +0.150 &  171(4) & 52(9) & 0.12(4) &    \le 0.63 \\
 -0.655       & +0.100 &  161(6) & 20(2) & 0.29(3) &     0.21(5) \\
 -0.862       & +0.100 &  152(1) & 18(1) & 0.54(4) &     0.24(2) \\
 -0.950       & +0.150 &  129(1) & 23(2) & 0.42(5) &     0.40(10) \\
 -0.982       & +0.100 &  130(2) & 18(2) & 0.30(3) &     0.27(5) \\
\hline
\multicolumn{6}{c}{\bf O} \\
 +2.895       & +0.050 &    8(2) & 24(4) & 0.30(2) & \le 0.38      \\
 +2.880       & +0.113 &    9(1) & 22(2) & 0.32(4) &     0.28(1)   \\
 +2.720       & -0.150 &    9(4) & 19(6) & 0.07(4) & \le 0.55 \\
 +2.610       & +0.000 &   -9(2) & 28(5) & 0.14(3) & \le 0.47 \\
 +2.610       & +0.000 &  -39(3) & 16(6) & 0.07(3) & \le 0.78 \\
 +2.568       & +0.150 &    4(2) & 23(4) & 0.12(3) & \le 0.60 \\
 +2.121       & -0.150 &  -47(1) & 13(3) & 0.14(3) & \le 0.37 \\
 +2.121       & -0.150 &   25(4) & 49(8) & 0.06(3) & \le 0.68 \\
 +2.056       & -0.300 &    1(2) & 23(2) & 0.17(3) & \le 0.31 \\
 +2.056       & -0.300 &   36(1) & 18(3) & 0.11(3) & \le 0.44 \\
\hline
\multicolumn{6}{c}{\bf   P}                \\
-0.590$^{a}$ & -0.050 &  -41(1) & 14(2) & 0.28(2) & \le 0.59      \\
-0.688$^{a}$ & -0.050 &  -41(1) & 26(3) & 0.42(4) & \le 0.47      \\
-0.688$^{a}$ & -0.050 &  -17(1) & 16(2) & 0.57(4) &     0.51(7)   \\
-0.688$^{a}$ & -0.050 &    2(1) & 25(2) & 0.59(4) &     0.47(7)   \\
-0.754       & -0.050 &  -10(2) & 37(3) & 0.32(4) &     0.15(4)   \\
-0.829       & -0.300 &  -18(1) & 40(2) & 0.22(3) & \le 0.35      \\
-0.906       & -0.150 &   -2(2) & 39(2) & 0.40(5) &     0.24(4)   \\
-0.655       & +0.100 &    2(2) & 24(7) & 0.09(3) & \le 0.35      \\
-0.862       & +0.100 &   -1(1) & 23(3) & 0.25(4) &     0.24(5)   \\
\hline
\multicolumn{6}{c}{\bf J}                \\
+2.720       & -0.150 &   96(1) & 23(3) & 0.29(4) &     0.45(8)   \\
+2.056       & -0.300 &  141(2) & 23(2) & 0.16(3) & \le 0.33      \\
+1.958$^{a}$ & -0.100 &  122(3) & 27(1) & 0.56(6) &     0.25(3)   \\
\hline
\multicolumn{6}{c}{\bf K}                \\
+1.958$^{a}$ & -0.100 &  247(1) &  9(3) & 0.22(6) & \le 0.32      \\
+1.708       & -0.150 &  221(4) & 31(6) & 0.08(2) & \le 0.85      \\
+1.698$^{a}$ & -0.150 &  220(3) & 20(3) & 0.16(3) & \le 0.20      \\
+1.522$^{a}$ & -0.150 &  219(3) & 16(1) & 0.42(4) & \le 0.09      \\
+1.403$^{a}$ & -0.150 &  214(6) & 21(5) & 0.14(3) &     0.14(4)   \\
+1.310$^{a}$ & -0.130 &  204(2) & 19(4) & 0.11(9) & \le 0.53      \\
+0.401       & -0.150 &  192(3) & 28(8) & 0.08(4) & \le 0.49      \\
+0.390$^{a}$ & -0.150 &  195(2) &  8(1) & 0.39(2) & \le 0.12      \\
+0.145$^{a}$ & -0.150 &  175(2) &  7(1) & 0.20(3) & \le 0.38      \\
+1.698$^{a}$ & -0.150 &  185(3) & 12(3) & 0.11(3) &     0.5(2)    \\
+1.698$^{a}$ & -0.150 &  163(3) & 13(3) & 0.12(3) & \le 0.25      \\
+1.708       & -0.150 &  157(6) & 48(9) & 0.53(2) & \le 0.01      \\
+0.227       & -0.050 &  176(1) & 10(1) & 0.20(5) & \le 0.39      \\
+0.150$^{a}$ & -0.050 &  154(1) &  4(1) & 0.20(6) & \le 0.42      \\
+0.145$^{a}$ & -0.080 &  177(2) & 12(4) & 0.21(9) & \le 0.47      \\
\hline
\end{tabular}
\end{center}
\end{table}

\addtocounter{table}{-1}
\begin{table}[p]
\caption{Continuation.}
\begin{center}
\begin{tabular}{llllll}
\hline
$l$ & $b$ & $v$  & $\Delta v$  & I(CS) & $\frac{I(SiO)}{I(CS)}$ \\
\hline
\multicolumn{6}{c}{\bf M}                    \\
 +0.401       & -0.150 & -102(4) & 53(8) & 0.10(4) & \le 0.42 \\
 +0.390$^{a}$ & -0.150 &  -92(2) & 12(4) & 0.06(2) &     0.7(3) \\
 +0.390$^{a}$ & -0.150 &  -71(3) & 18(8) & 0.05(2) &     1.0(5) \\
 +0.145$^{a}$ & -0.150 & -131(1) &  8(3) & 0.09(3) & \le 0.71 \\
 +0.145$^{a}$ & -0.150 & -107(5) & 64(9) & 0.10(3) & \le 0.66 \\
 +0.063       & -0.300 & -140(2) & 38(4) & 0.12(3) & \le 0.40 \\
 -0.460       & -0.300 & -118(1) & 22(1) & 0.60(4) &     0.09(2) \\ 
 -0.829       & -0.300 & -123(3) & 39(9) & 0.10(3) & \le 0.66 \\
 -0.906       & -0.150 & -133(1) & 22(3) & 0.19(5) & \le 0.27 \\
 -0.906       & -0.150 & -169(2) & 27(8) & 0.11(5) & \le 0.40 \\
 +0.945$^{a}$ & -0.050 &  -81(2) & 14(3) & 0.23(8) &     0.3(2) \\
 +0.945$^{a}$ & -0.050 &  -65(2) & 14(3) & 0.22(8) &     0.18(9) \\
\hline
\multicolumn{6}{c}{\bf   N}            \\
-1.102       & -0.250 &  -44(2) & 39(4) & 0.21(5) &     0.26(11)  \\
-1.189       & -0.300 &   -4(1) & 13(4) & 0.19(4) & \le 0.58      \\
-1.189       & -0.300 &  -42(1) & 41(3) & 0.43(4) &     0.21(5)   \\
-1.189       & -0.300 & -103(3) & 15(6) & 0.08(4) & \le 1.08 \\
-1.211       & -0.150 & -118(2) & 25(2) & 0.50(4) &     0.26(6) \\
-1.211       & -0.150 &  -42(1) & 34(2) & 0.46(4) &     0.28(6)   \\
\hline
\end{tabular}
\end{center}
\end{table}

\bibliographystyle{aa}
\bibliography{nms}

\end{document}